\documentclass{ws-procs975x65}

\begin{document}



\title{COMPACT BINARY COALESCENCE AND THE SCIENCE CASE FOR EINSTEIN TELESCOPE}

\author{CHRIS VAN DEN BROECK}

\address{
Nikhef, Science Park 105, 1098 XG Amsterdam, The Netherlands\\
\email{vdbroeck@nikhef.nl}}

\bodymatter

\vskip 0.1cm

\emph{Einstein Telescope.} A design study for a 3rd generation ground-based gravitational wave detector has been funded under the European Framework Programme-7. The basic design of Einstein Telescope (ET) is still under discussion [\citen{Freiseetal}], but one possibility is to have three interferometers with a $60^\circ$ opening angle and 10 km arm length, arranged in an equilateral triangle. A single ET interferometer is envisaged to be sensitive between 1 Hz and several kHz, with a strain sensitivity that is roughly an order of magnitude better than that of Advanced LIGO. This will allow it to see binary neutron star coalescences out to $z \sim 2$, and binary black holes out to $z \sim 8$ . A straightforward extrapolation from the expected Advanced LIGO detection rate suggests that ET will see in the order of $10^6$ binary coalescence signals per year.

\begin{figure}[ht]
\begin{center}$
\begin{array}{cc}
\includegraphics[height=4cm]{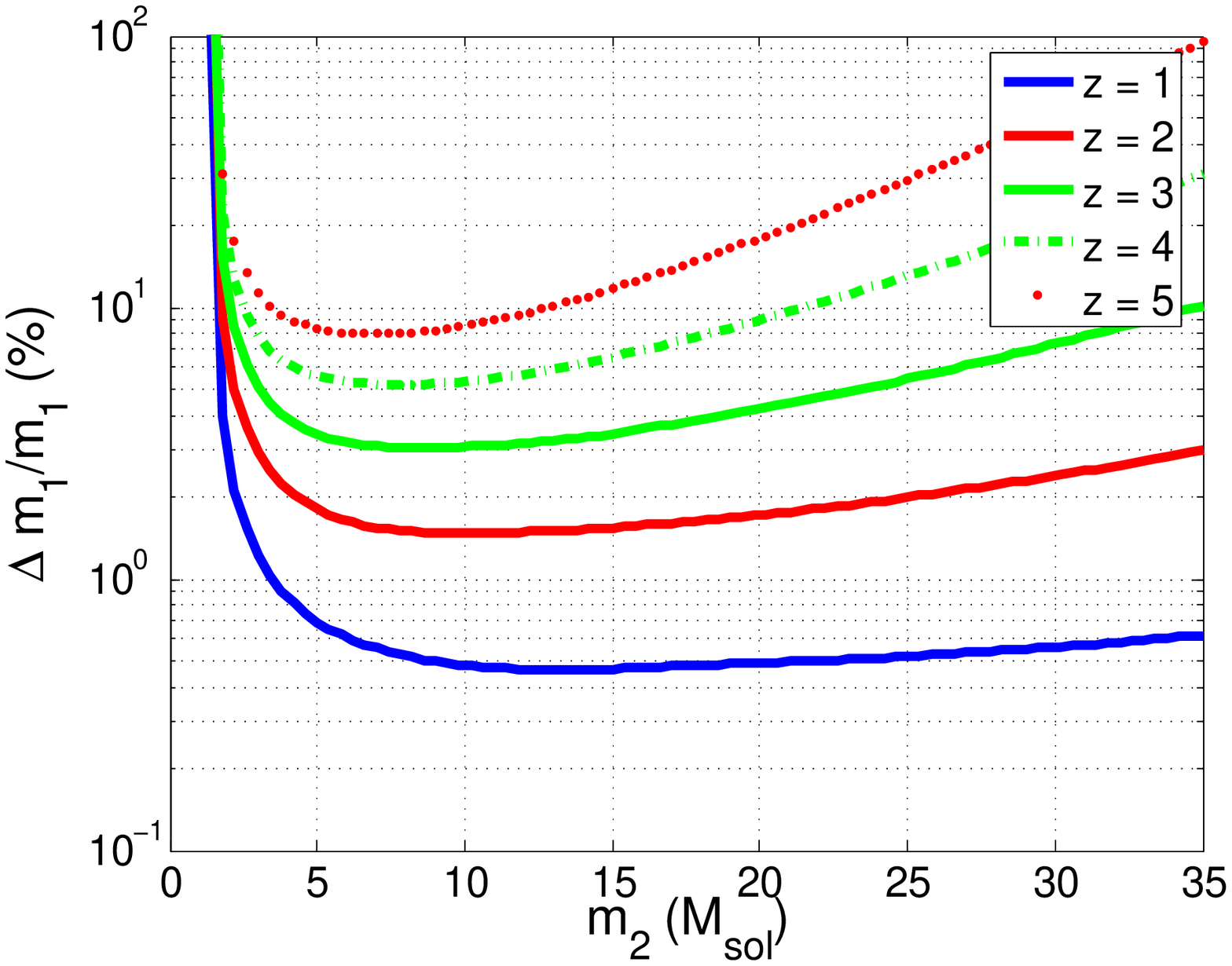}&
\includegraphics[height=4cm]{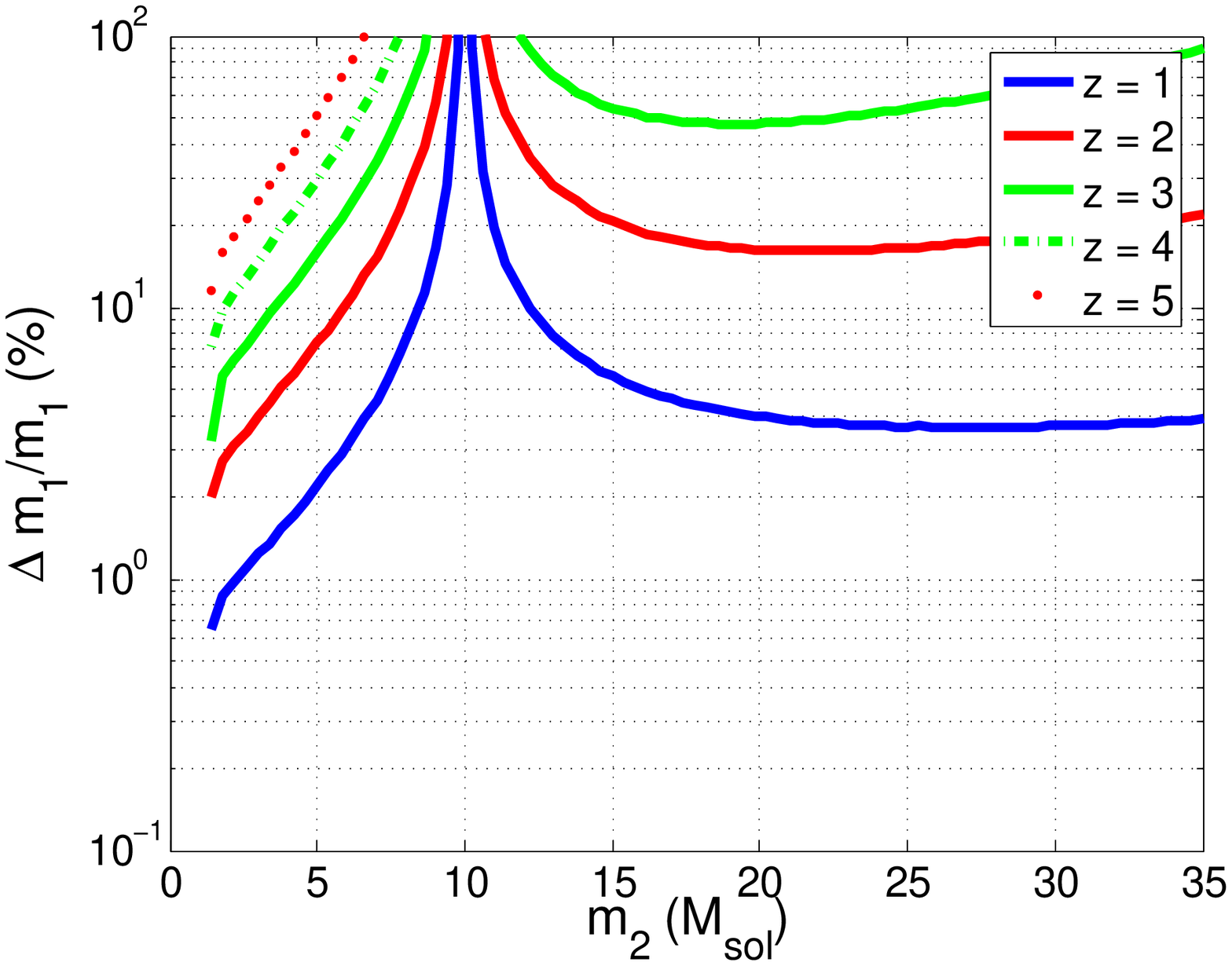}
\end{array}$
\end{center}
\caption{The angle-averaged accuracy with which ET can measure the mass of a $1.4\,M_\odot$ neutron star (left) and a $10\,M_\odot$ black hole in NS-BH inspirals, as a function of the mass of the secondary object, and depending on the redshift.}
\label{fig:NSBHmass}
\end{figure}

\emph{The mass function of neutron stars and black holes.} Given this enormous number of sources, ET will be able to make a census of the mass distributions 
of neutron stars and black holes, and their time evolution, over cosmological timescales. Fig.~\ref{fig:NSBHmass} shows the uncertainty (averaged over 
sky position and orientation of the orbital plane) with which the mass of a neutron star or a black hole in a binary neutron star (BNS), a neutron star-black hole (NS-BH) or binary black hole (BBH) event can be measured from the inspiral waveform alone. In the equal mass case it will be difficult to extract the mass, but as we will discuss in the next section, the merger will provide the missing information at least for neutron stars. Models of the evolution of star formation rate suggest that the rate at which coalescences occur will peak at $z \simeq 1 - 3$ [\citen{rates}], a range where ET can measure masses with high accuracy. Thus, ET will be able to make a detailed map of how neutron star and black hole masses evolved during the most interesting part of star formation history.

\emph{Internal structure of neutron stars.}
An important open question is the behavior of the kind of dense nuclear matter in a neutron star, encapsulated by the equation of state (EOS).
When the separation of the compact objects becomes of the same order as their size, neutron stars are tidally deformed, which in turn affects the orbital motion, and hence the gravitational waveform [\citen{Hinderer:2009}]. The radius $R$ and pressure $p$ of neutron stars can be determined from finite size effects during the late stages of inspiral; at 100 Mpc, $\Delta R/R \sim 0.05 - 0.1$, and about the same range for $\Delta p/p$. Especially the shape of the merger signal is quite sensitive to the EOS; for a representative example, see [\citen{Baiotti:2008}]. Since the difference in mass choices considered by the latter authors was 10\%, and given the large SNRs (hundreds up to 1 Gpc), ET should be able to infer masses to at least that accuracy.   

\begin{figure}[ht]
\begin{center}$
\begin{array}{cc}
\includegraphics[height=4cm]{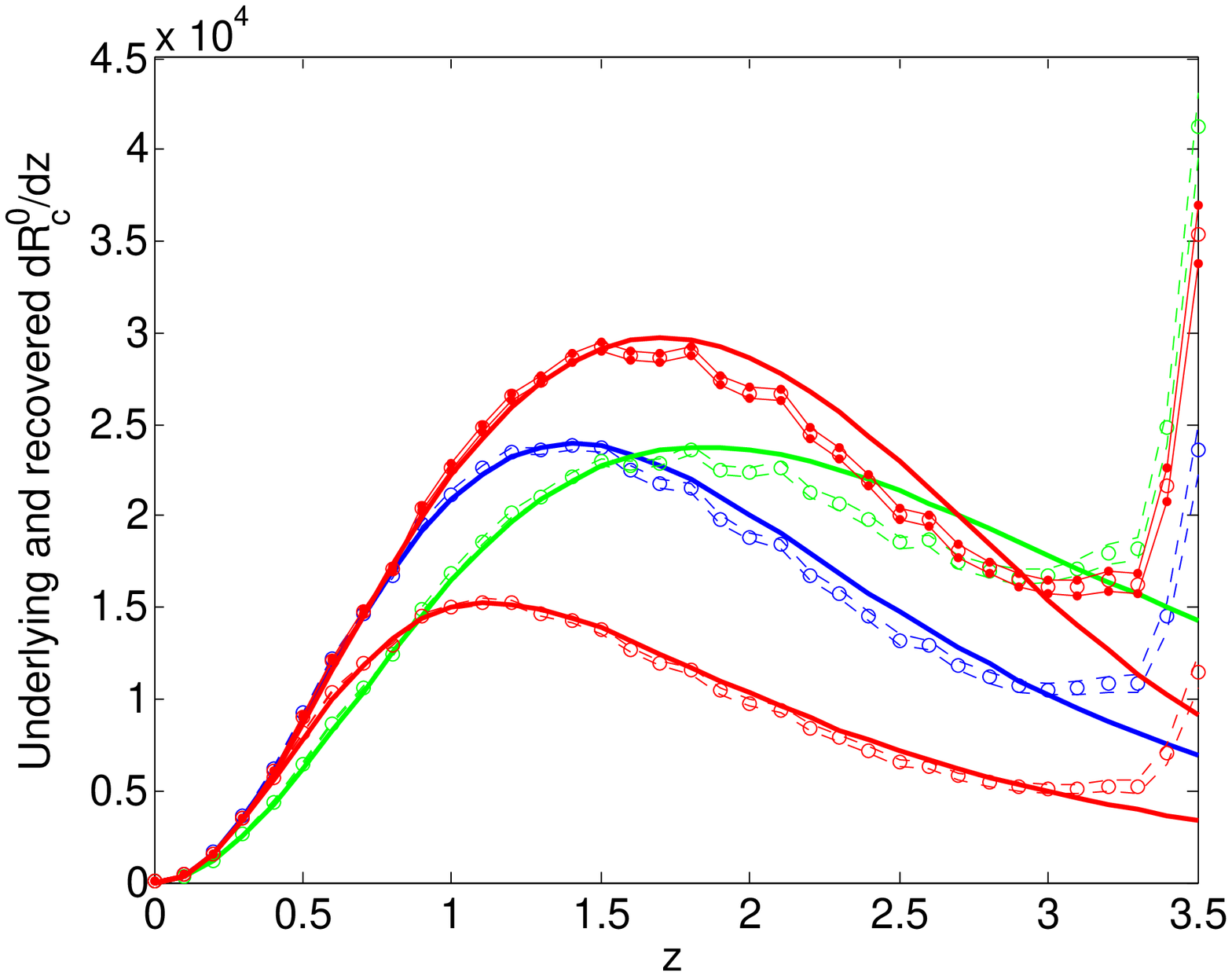}&
\includegraphics[height=4cm]{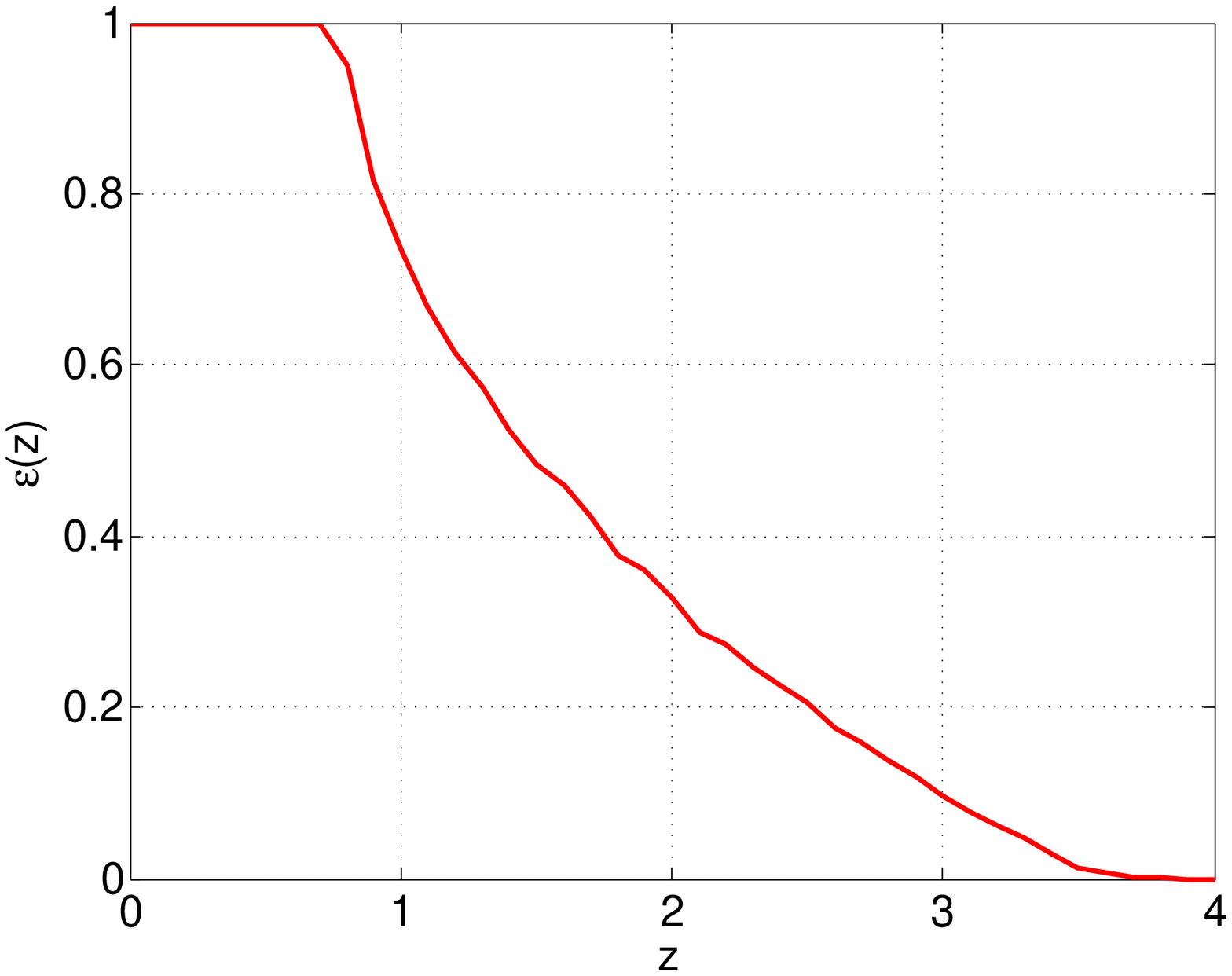}
\end{array}$
\end{center}
\caption{Left: Underlying and recovered coalescence rates for the models of Hopkins and Beacom (top, red), Fardal et al.~ (blue), Wilkins et al.~(bottom, red), and Nagamine et al.~(green) [\citen{rates}]. The solid lines are the true rates, the circles give the number of measured coalescences in a redshift bin, and the dashed lines give a 1-sigma spread in recovered redshifts. The recovered rates take into account loss of detection efficiency, shown on the right.}
\label{fig:ratesrecovery}
\end{figure}

\emph{Coalescence rates.} By measuring the luminosity distance from the signal, and using a cosmological model to convert to redshift, one can estimate the coalescence rate $dR^0_c/dz(z)$. Fig.~\ref{fig:ratesrecovery} illustrates this for four models in the recent literature. Errors in the measured distances due to both detector noise and weak lensing effects cause events to be erroneously binned in redshift. The recovered rate also needs to be corrected for loss of detection efficiency, causing it to eventually diverge as the efficiency goes to zero. However, up to $z \lesssim 1.5$ the error is in the order of 1\%.  

\begin{figure}[ht]
\begin{center}$
\begin{array}{ccc}
\includegraphics[height=4cm]{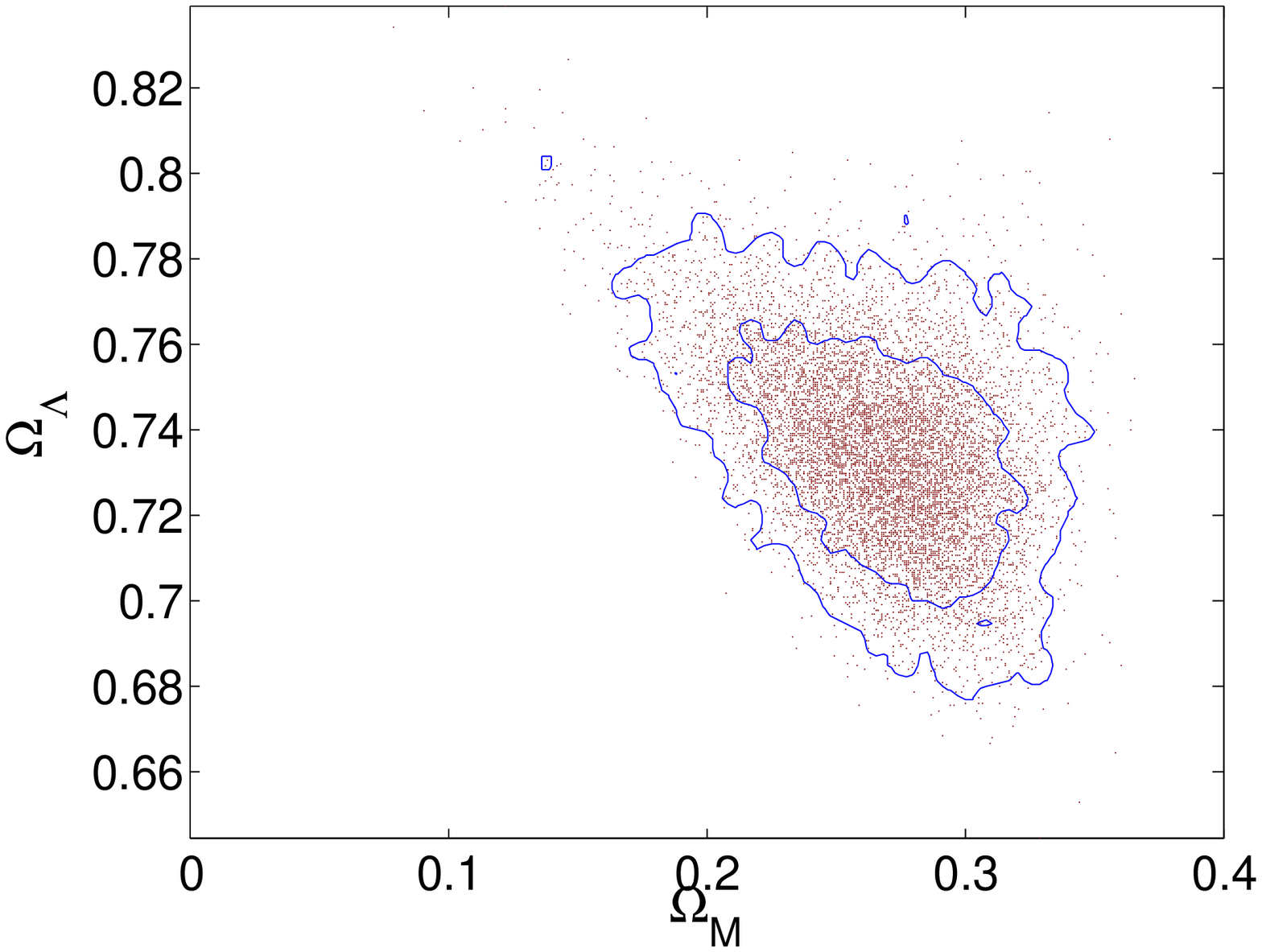}&
\includegraphics[height=4cm]{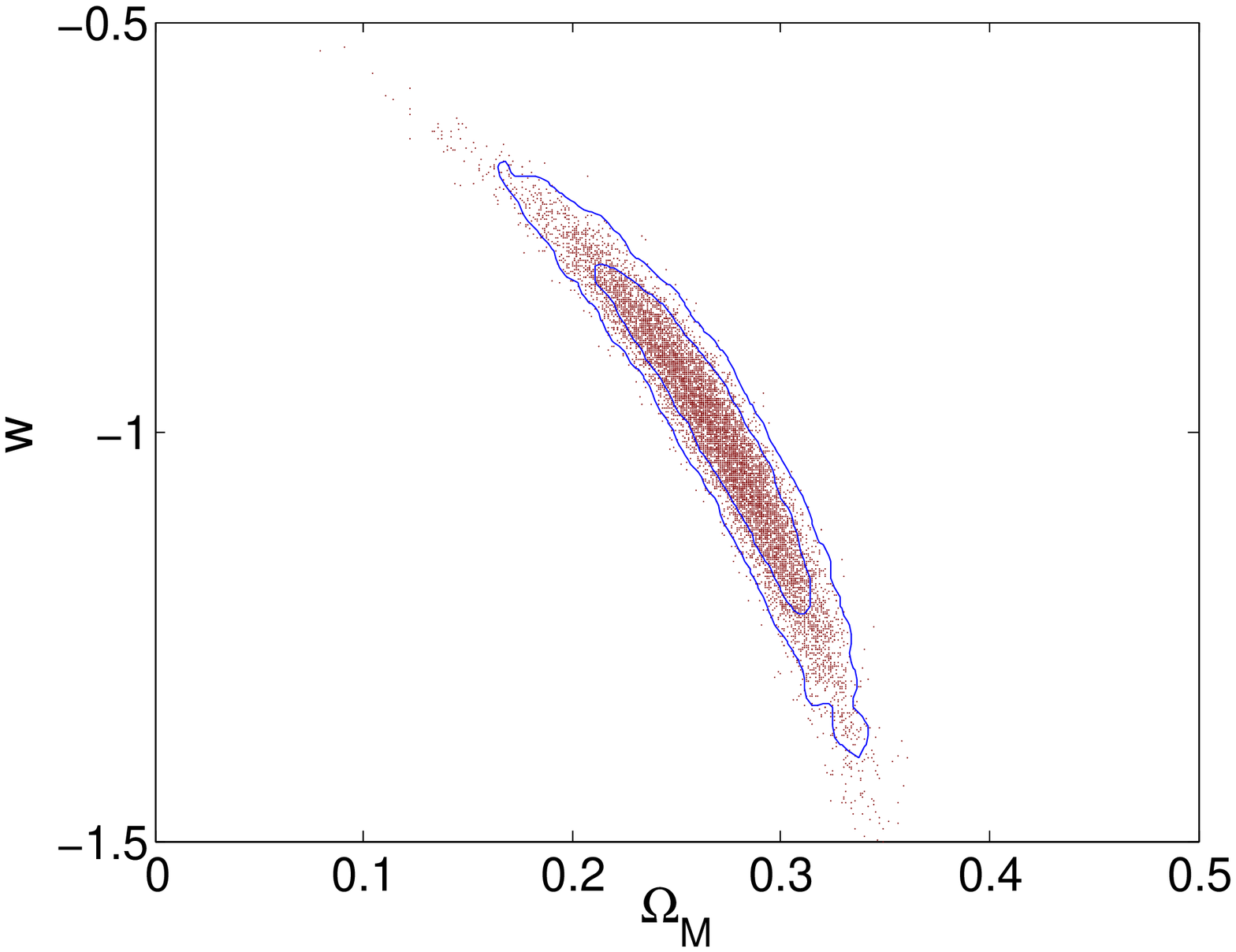}
\end{array}$
\end{center}
\caption{Scatter plots for the values of $\Omega_{\rm M}$, $\Omega_{\rm \Lambda}$, and $w$ found by fitting $D_{\rm L}$ against $z$ for 10,000 ``catalogs" of $\sim$1,000 sources each. The solid lines are 1-sigma and 2-sigma contours. One has $\sigma_{\Omega_{\rm M}}/\Omega_{\rm M} = 12.7\%$, $\sigma_{\Omega_{\rm \Lambda}}/\Omega_{\rm \Lambda} = 3.0\%$, and $\sigma_w/|w| = 14.5\%$. (These numbers differ sligthly from those in [\citen{Sathya:2009}]; here we used a more realistic coalescence rate distribution [\citen{rates}].)}
\label{fig:cosmography}
\end{figure}

\emph{Cosmography.} Coalescences involving at least one neutron star will have electromagnetic (EM) counterparts (e.g., gamma ray bursts) and can be used as ``standard sirens" [\citen{Schutz:1986}]. From the gravitational waveform one can extract the luminosity distance $D_{\rm L}$, while identification of the host galaxy through the EM counterpart yields redshift $z$. Assuming that the host can be identified for $\sim 1000$ binary neutron star coalescences over several years, one can fit the function $D_{\rm L}(z)$, which contains information about, e.g., the matter density $\Omega_{\rm M}$, dark energy density $\Omega_{\rm \Lambda}$, and equation-of-state parameter $w$. Uncertainties in the parameters will be determined by uncertainties in $D_{\rm L}$ due to detector noise and weak lensing. Fig.~\ref{fig:cosmography} shows parameter distributions from 10,000 simulated ``catalogs" of sources; see [\citen{Sathya:2009}] for details of the method. While ET observations will not be better than conventional methods, gravitational wave ``sirens" are self-calibrating and do not have to rely on the lower rungs of the cosmic distance ladder.

\section*{Acknowledgements}

It is a pleasure to thank R.~O'Shaughnessy, J.S.~Read, T.~Regimbau, L.~Rezzolla, B.S.~Sathyaprakash, B.~Schutz, and all the members of the ET Astrophysics Working Group for very useful discussions. This research was supported in part by STFC grant PP/F001096/1 and by the research programme of the Foundation for Fundamental Research on Matter (FOM), which is partially supported by the Netherlands Organisation for Scientific Research (NWO).


\begin{thebibliography}{00}

\bibitem{Freiseetal} A.~Freise, S.~Chelkowski, S.~Hild, W.~Del Pozzo, A.~Perecca, and A.~Vecchio (2008); S.~Hild, S.~Chelkowski, and A.~Freise (2008); S.~Hild, S.~Chelkowski, A.~Freise, J.~Franc, N.~Morgado, and R.~DeSalvo (2009); A.~Freise, S.~Hild, K.~Somiya, K.~Strain, A.~Vicer\`{e}, M.~Barsuglia, and S.~Chelkowski (2009).

\bibitem{rates} A.~Hopkins and J.~Beacom, Astrophys.~J.~{\bf 651}, 142 (2006); K.~Nagamine, J.~Ostriker, M.~Fukugita, and R.~Cen, Astrophys.~J.~{\bf 653}, 881 (2006); M.~Fardal, N.~Katz, D.~Weinberg, and R.~Dav\'{e}, MNRAS {\bf 379}, 985 (2007); S.~Wilkins, N.~Trentham, and A.~Hopkins, MNRAS {\bf 379}, 985 (2007).

\bibitem{Hinderer:2009} T.~Hinderer, B.D.~Lackey, R.N.~Lang, and J.S.~Read (2009).

\bibitem{Baiotti:2008} L.~Baiotti, B.~Giacomazzo, and L.~Rezzolla, Phys.~Rev.~D {\bf 78}, 064054 (2008).

\bibitem{Schutz:1986} B.~Schutz, Nature {\bf 323}, 310 (1986).

\bibitem{Sathya:2009} B.S.~Sathyaprakash, B.~Schutz, and C.~Van Den Broeck (2009), arXiv:0906.4151.






\end{thebibliography}
\end{document}